# ON Λ AND Ω MEASUREMENTS AND ACCELERATION OF UNIVERSE EXPANSION

O.G. Semyonov[1]

*State University of New York at Stony Brook, Stony Brook, NY 11794, USA*

**Abstract:** Luminosity distances to the high-redshift Ia supernovas measured by Supernova Cosmology Project team with the relativistic correction factors for redshift and time dilation led to the conclusion about acceleration of universe expansion. The effect of relativistic aberration of high-redshift sources requires an additional correcting factor for the observed luminosities of SN Ia supernovas which can reduce the data to fit the universe with Λ = 0, i.e. without acceleration.

## 1. INTRODUCTION

In their original publications (Perlmutter at al. 1998, 1999), the team of Supernova Cosmology Project calculated the luminosity distance $D_L$ to the high-redshift supernovas SN Ia (Perlmuter et al. 1998) from the measured luminosities of high-z SN with the correction factors for time dilation and redshift to scale the data with the known luminosities of low-z supernovas. The best fit (Perlmutter et al. 1999) of corrected magnitudes $m_{eff}$ corresponds to $\Omega_M = 0.28$, $\Omega_\Lambda = 0.72$ with a cosmological constant $\Lambda \neq 0$ leading to conclusion about the acceleration of universe expansion.

## 2. LUMINOSITY DIMMING DUE TO RELATIVISTIC ABERRAION

Beside the time dilation and wavelength shift, the effect of light aberration of relativistic sources with respect to an observer at rest (Møller 1972) also influences the flux measured by a detector. Consider two coordinate systems S and S´ both originated from an isotropic source at an arbitrary moment of time as shown in Fig. 1. The system S is at rest with respect to a detector and the system S´ co-moves together with the source with the velocity β along z-axis in positive direction. A light beam that makes an angle θ´ with the velocity vector (z-axis) in the frame S´ co-moving with the source will have apparent direction in the reference frame S characterized by another angle θ, where θ´ and θ are linked by:

$$\cos\theta' = \frac{\cos\theta - \beta}{1 - \beta\cos\theta}$$

with the azimuth angles φ = φ´ in the corresponding spherical coordinates. This formula displays, in particular, the well-known searchlight effect or the beaming of radiation in the direction of movement when it is observed from the reference (detector) unmoving frame (for example, θ = arccos β < π/2 when θ´ = π/2). It can be said, that from the point of view of an observer in the frame S, the solid angles made by the beams emitted from the source are compressed with respect to the co-moving coordinate system in the hemisphere turned towards the direction of movement and stretched out in the opposite hemisphere. Such the beaming is the well-known phenomenon in physics; it is observed, in particular, for synchrotron radiation of relativistic electrons moving in magnetic fields, for impact bremsstrahlung of relativistic electrons incident on a target at rest, and for the products of nuclear reactions of relativistic particles with particles at rest. Resent observations of jets emitted by AGN also displayed the strong evidence of beaming of radiation from the jets (Biretta, Sparks, & Macchetta 1999) and superluminosity/dimming of incoming/receding jets.

---

[1] osemyonov@ece.sunysb.edu

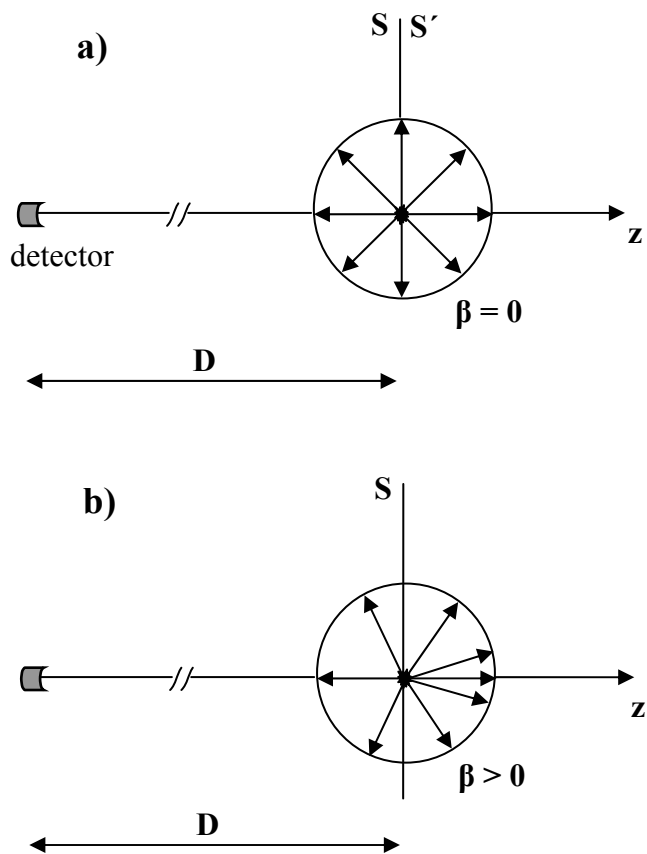

Fig. 1: a) Angular distribution of light rays emitted by an immobile isotropic source in the coordinate frame S related to a detector and by a moving source in the co-moving coordinate frame S´; b) angular distribution of the same rays for the moving relativistic source as seen from the immobile coordinate frame S: the rays in the right hemisphere are inclined toward the z-axis (direction of movement) compressing the solid angles made by corresponding light rays and boosting the source luminosity (number of photons per steradian), while the rays in the left hemisphere decline out of z-axis stretching the solid angles and diminishing the source luminosity in the direction to the detector. D is the actual distance from the detector to the source in the frame at rest, which is smaller then the luminosity distance $D_L$ evaluated from light flux

measurements. The sketch is drafted to demonstrate the phenomenon of luminosity change in the frame at rest due to relativistic aberration and does not reflect other relativistic effects.

It can be shown from the equation above (Kraus 2000; McKinley 1980), that a backward-directed small solid angle made by the light rays originated from a source moving away from a detector along the line of sight in a flat space is transformed with respect to the co-moving coordinate system as $d\Omega = \delta^2 d\Omega' = d\Omega'/[\gamma^2(1-\beta)^2] = d\Omega' (1+z)^2$. The effect results in dimming of relativistic source luminosity by a factor of $(1+z)^{-2}$ because of the diminished flux $F$ on the detector from the relativistic source with luminosity $L$ in the co-moving frame due to decreased number of detected photons per unit area at any distance from the source in the coordinate frame S and, therefore, in apparent luminosity distance $D_L = (L/4\pi F)^{1/2}$ increase by the factor of $(1+z)$. In general, the spectrally integrated flux is transformed as $F = F'(1+z)^{-4}$ when the correcting factors for time dilation and redshift are taken into account together with the effect of solid angle widening (Kraus 2000). The same transformation rule with the transforming factor of $(1+z)^{-4}$ for the spectrally integrated radiation flux can also be obtained from the relativistic invariant $L_\nu/\nu^3$ ($\nu$ is the radiation frequency and $L_\nu$ is the spectral luminosity) allowing for time stretching. Therefore, to scale the observation data for the high-redshift supernovas in the template for the low-z supernovas, an additional negative correction term should be added to the expression for the calculated magnitudes $m_B^{eff}$ equal to $\Delta m = -5\log(1+z)$ (e.g., $\Delta m \approx -0.88$ mag for z = 0.5). This factor seems sufficient to fit the universe with the cosmological constant $\Lambda$ close to zero within the errors of measurements, i.e. the universe without acceleration.